\documentclass{article}

\PassOptionsToPackage{numbers, compress, square}{natbib}

\usepackage[final]{neurips_2019_ml4ps}

\usepackage[utf8]{inputenc} % allow utf-8 input
\usepackage[T1]{fontenc}    % use 8-bit T1 fonts
\usepackage{hyperref}       % hyperlinks
\usepackage{url}            % simple URL typesetting
\usepackage{booktabs}       % professional-quality tables
\usepackage{amsfonts}       % blackboard math symbols
\usepackage{nicefrac}       % compact symbols for 1/2, etc.
\usepackage{microtype}      % microtypography
\usepackage[pdftex]{graphicx}

\usepackage{caption} 
\newcommand\solphys{\ref@jnl{Sol.~Phys.}}% Solar Physics 

\title{Auto-Calibration of Remote Sensing Solar Telescopes with Deep Learning}
\author{
    Brad Neuberg\\
  NASA Frontier Development Lab\\
  \And
  Souvik Bose\\
  Rosseland Center for Solar Physics, \\ University of Oslo
  \And
  Valentina Salvatelli  \\
  NASA Frontier Development Lab \& IQVIA\\
  \And
  Luiz F.G. dos Santos\\
  CUA-IACS, NASA-GSFC\\
  \AND
  Mark Cheung\\
  Lockheed Martin Solar and Astrophysics Laboratory \& \\
  Stanford University \\
  \And
  Miho Janvier\\
  Institut d'Astrophysique Spatiale, Université Paris-Sud\\
  \And
  Atilim Gunes Baydin\\
  University of Oxford\\
  \And
  Yarin Gal \\
  OATML, University of Oxford \\
  \And
  Meng Jin\\
  Lockheed Martin Solar and Astrophysics Laboratory \& SETI\\
  }

\begin{document}

\maketitle

\begin{abstract}
  As a part of NASA's Heliophysics System Observatory (HSO) fleet of satellites, the Solar Dynamics Observatory (SDO) has continuously monitored the Sun since 2010. Ultraviolet (UV) and Extreme UV (EUV) instruments in orbit, such as SDO's Atmospheric Imaging Assembly (AIA) instrument, suffer time-dependent degradation which reduces instrument sensitivity. Accurate calibration for (E)UV instruments currently depends on periodic sounding rockets, which are infrequent and not practical for heliophysics missions in deep space. In the present work, we develop a Convolutional Neural Network (CNN) that auto-calibrates SDO/AIA channels and corrects sensitivity degradation by exploiting spatial patterns in multi-wavelength observations to arrive at a self-calibration of (E)UV imaging instruments. Our results remove a major impediment to developing future HSO missions of the same scientific caliber as SDO but in deep space, able to observe the Sun from more vantage points than just SDO's current geosynchronous orbit. This approach can be adopted to perform autocalibration of other imaging systems exhibiting similar forms of degradation.
\end{abstract}

\section{Introduction}
\label{intro}
Solar activity plays a major role in influencing the interplanetary medium and space-weather around us. Understanding the complex mechanisms that govern such a dynamic phenomenon is important and challenging. Remote-sensing instruments on board heliophysics missions can provide a wealth of information on the Sun's activity, especially via the measurement of magnetic fields and the emission of light from the multi-layered solar atmosphere. NASA currently operates the Heliophysics System Observatory (HSO) that consists of a fleet of satellites constantly monitoring the Sun, its extended atmosphere and space environments around the Earth and other planets of the solar system \citep{HSO}.

One of the flagship missions of the HSO is NASA’s Solar Dynamics Observatory (SDO) \citep{SDO_primary}. Launched in 2010, it consists of three instruments: the Atmospheric Imaging Assembly (AIA) \citep{AIA}, the Helioseismic \& Magnetic Imager (HMI) \citep{HMI}, and the EUV Variability Experiment (EVE) \citep{EVE}. The SDO has been generating terabytes of observational data every day and has been constantly monitoring the Sun with the highest temporal and spatial resolution for full-disk observations.

Unfortunately, the (E)UV instruments in orbit suffer time-dependent degradation which reduces instrument sensitivity \citep{AIA_calib_paper}. Accurate calibration for EUV instruments currently depends on sounding rockets (e.g., for SDO/EVE, and SDO/AIA), that are infrequent \citep{EVE_rocket}. Since SDO is in a geosynchronous orbit, sounding rockets can be used for calibration, but calibration experiments may not be practical for missions in deep space (e.g., STEREO satellites) \citep{STEREO}.

In the present work, we develop a neural network that auto-calibrates the SDO/AIA channels, correcting sensitivity degradation, by exploiting spatial patterns in multi-wavelength observations to arrive at a self-calibration of (E)UV imaging instruments. This removes a major impediment to developing future HSO missions that can deliver solar observations from different vantage points beyond Earth-orbit.

\section{Methodology}
\label{method}

\subsection{Dataset}
The processed SDO AIA dataset \citep[][hereafter SDOML]{SDOML} was used in order to provide a machine-learning ready dataset for the auto-calibration problem. More specifically, the dataset, which consists of a subset of the original SDO data dating from 2010 to 2018, is comprised of 7 EUV channels, 2 UV channels, and HMI vector magnetograms (three components). Both AIA and HMI data are spatially co-registered, have identical spatial sampling, and all instruments are synchronous. The temporal cadence is 6 min for AIA, 12 min for HMI, while the spatial resolution is \textasciitilde{}4.8 arcsec per pixel (each full disk image has 512 x 512 pixels). Crucially, the images are corrected for instrumental degradation over time and have exposure corrections (determined from sounding rocket experiments). The end result is that in this dataset, changes in pixel brightness are from changes in solar conditions and not due to instrument performance.

\begin{figure}[!htb]
	\centering
	\hspace*{-0.4in}%
	\begin{minipage}{0.4\textwidth}
		\centering
        \includegraphics[height=2.3in]{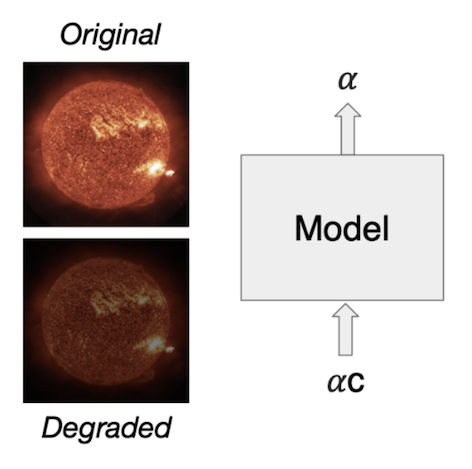}
        \caption{A schematic of the auto-calibration problem. $\alpha$ is the brightness degradation factor.}
        \label{fig:autocalibrate_model_problem}
	\end{minipage}%
    \hspace{19mm}%
	\begin{minipage}{0.4\textwidth}
		\centering
		\hspace*{-0.45in}
        \includegraphics[height=2.3in]{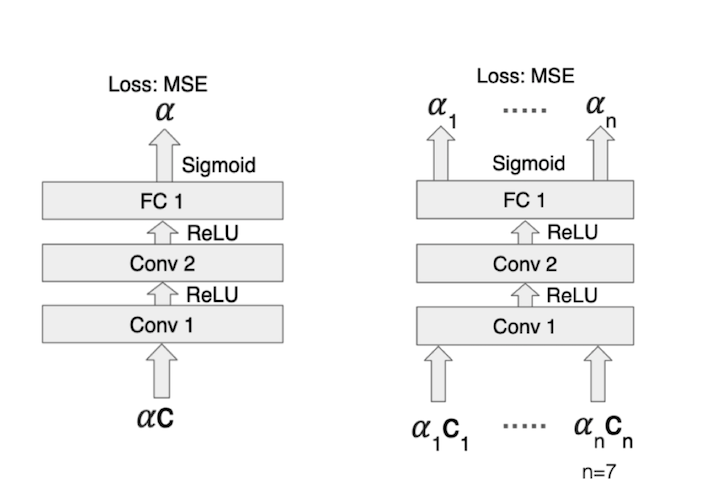}
        \caption{CNN taking in one channel (left) and multiple channels (right).}
        \label{fig:autocalibrate_CNN_arch}
	\end{minipage}
\end{figure}

\subsection{Statement of Problem}
Consider a set $\mathbf{C} = \{C_i, i\in [1,...,n]\}$ of multi-channel, contemporaneous set of SDO/AIA images sampled from the SDOML dataset.  $C_i$ denotes the $i$-th channel image in the set. Consider a corresponding set of dimmed images $\bold{\alpha} \mathbf{C} = \{\alpha_i C_i, i\in [1,...,n]\}$, where each dimming factor $\alpha_i$ is individually sampled from a uniform distribution [$0.01, 1.0$]. Note the dimming factors $\alpha_i$ apply uniformly per channel and is not spatially dependent. The spatial dependence portion of the degradation is assumed to be accounted for by periodic flatfields applied to AIA images. The goal is to find a model $M:  \bold{\alpha} \mathbf{C} \rightarrow \mathbf{\alpha}$, i.e. find a model that retrieves the set of multi-channel dimming factors $\alpha_i$ (see Fig. \ref{fig:autocalibrate_model_problem}).

In actuality, the realized dimming factors $\alpha_i(t)$ (where $t$ is time since beginning of the SDO mission) trace a single trajectory in $n$-dimensional space (starting with $\alpha_i(t=0) = 1 \forall ~i\in[1,...,n]$ at the beginning of the mission).
During model training, we intentionally hide this time-dependence from the model in order to make it more general and robust. This is done by (1)  using the SDOML dataset which has already been corrected for degradation effects, and (2) not assuming any relation between $t$ and $\bold{\alpha}$ and not using $t$ as an input feature.

The training set comprises multi-channel images $\mathbf{C}$ taken during the months January to July during the years 2012 to 2013 (2 years). From one training epoch to another, the same multi-channel image set $\mathbf{C}$ can be dimmed by a completely different set of $\bold{\alpha}$ dimming factors. This is a data augmentation strategy that allows the model to generalize and be effective in retrieving dimming factors over a range of solar conditions.

The test set comprises images taken during the months August to October during the years 2012 to 2013. The outcome of the learning algorithm is tested via a metric, the “binary frequency”. This metric represents the percentage of predictions matching ground truth with an error less than a given tolerance across all the channels. We chose a binary frequency tolerance of 0.05 to gauge model success, which is below the uncertainty of the AIA degradation calibration \citep{AIA_calib_paper}.

\subsection{Baseline Model}
In order to compare the results of our deep learning approach, we created the following baseline model, which is motivated by the assumption that the EUV brightness outside magnetically active regions is invariant in time.

From the SDOML dataset, we picked a set of reference images per channel $C_{\rm ref}$ at time $t=t_{\rm ref}$. We computed the histograms (probability density functions) of the pixel values per channel. However, since the level of solar activity (in terms of active regions covering the solar disk) continuously evolves in time, so does the computed histograms. We propose that by including only pixels where the HMI line-of-sight magnetic flux density is $B_{\rm los} < 5$ Mx cm$^{-2}$, where Mx cm$^{-2}$ is a unit of magnetic flux density, the histograms would capture only information about the Sun's ambient background state (the so-called "quiet Sun"). For each AIA channel, the reference histogram has a maximum $I_{i,{\rm ref}}^{\rm mp}$ corresponding to the most probable brightness value. 

The baseline model works as follows. For any multi-channel image $\mathbf{C}$, we extract the channel-wise most probable value $I^{\rm mp}_{i}$. The baseline model estimates dimming factors according to: 
\begin{equation}
    \alpha_i := I^{\rm mp}_{i} / I_{i, {\rm ref}}^{\rm mp}.
\end{equation}

\noindent In terms of the binary frequency metric, the best results for the dimming factor retrieved from the ratio of the most probable values are 90\% (304 \AA) and 76\% (131 \AA), and we find \textasciitilde{}54\% mean success across all channels, for a tolerance level (the absolute difference between the ground truth degradation factor and the predicted one) of 0.05.

\section{Convolutional Neural Network Model}
\begin{table}[tb]
  \centering
  \bgroup
  \def\arraystretch{1.5}
  \begin{tabular}{|l|c|}
     \hline
     Model & Binary Frequency Correct (tolerance=0.05) \\
     \hline
     Mean baseline          & 54\%  \\
     \hline
     Mean single channel CNN    & 93\% \\
     \hline
     Worst single channel CNN   & 75\% \\
     \hline
     \textbf{Multi-channel CNN} & \textbf{100\%} \\
     \hline
  \end{tabular}
  \egroup
  \caption{Comparison between baseline, single channel Convolutional Neural Network (CNN) model (both mean and the worst wavelength), and multi-channel CNN model over 693 hold out test images.}
  \label{tab:autocalibrate_final_results}
\end{table}

As a first step, we tried a Convolutional Neural Network (CNN) \citep{6795724} taking in one AIA wavelength channel, as represented in the schematic on the left of Fig.~\ref{fig:autocalibrate_CNN_arch}.

The architecture of the CNN consists of two convolutional layers and one fully connected (FC) layer, as well as two ReLUs (Rectified Linear Units) \citep{Nair:2010:RLU:3104322.3104425} and one final sigmoid as the activation functions. We use a L2 loss function, measuring the mean squared error of the distance between the predicted degradation factor and the ground truth value. The software library PyTorch was used for both training and inference of the CNN \citep{paszke2017automatic}.

The results are given in Table~\ref{tab:autocalibrate_final_results}, showing that for a tolerance level of 0.05, using a single channel increases the metric from 54\% for the baseline to 93\% for the CNN (taking the mean of the metric across all wavelength channels). Furthermore, the metric for the worst wavelength degradation recovery (for 211 \AA) is at 75\%, which is still significantly better than the baseline.

\begin{figure}
  \centering
  \captionsetup{width=0.6\linewidth}
  \includegraphics[width=0.6\linewidth]{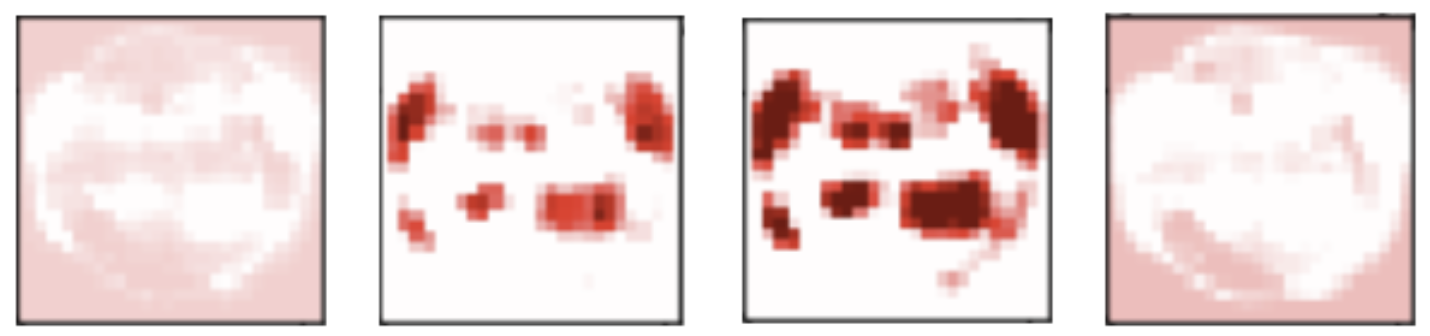}
  \caption{Subset of activations from the final convolutional layer, indicating that the convolutional layers have learned to distinguish between different morphological structures on the Sun.}
  \label{fig:autocalibrate_activation_viz}
\end{figure}

We also test the same CNN taking as input degraded images from multiple channels, as indicated in the schematic on the right of Fig.~\ref{fig:autocalibrate_CNN_arch}. The hypothesis tested here is that the multi-channel CNN can learn more effectively than a single channel CNN by taking into account structures appearing in the different channels. A visualization of what structures are learned by the final convolutional layer, produced via \citep{uozbulak_pytorch_vis_2019}, can be seen in Fig.~\ref{fig:autocalibrate_activation_viz}. Such a mapping allows to see that the network learns both to distinguish the structures of the Sun and to separate them from the background. The results are presented in Table \ref{tab:autocalibrate_final_results}: we can see that the multi-channel CNN recovers the degradation factor, for all channels and for a tolerance level of 0.05, to 100\%, over 693 test images. This is better than both the baseline as well as the CNN architecture that uses only one channel.

\subsection{Dependence on Inter-Channel and Morphology/Brightness Relationships}

In setting out to solve this problem, our premise was that the relationships between the brightness and morphology of solar features across multiple channels would facilitate the model to accurately estimate dimming factors. 

To test this premise, we generated  synthetic, idealized fake "Sun" images, in which the Sun is replaced by a 2D Gaussian profile of height $A$ and characteristic width $\sigma$, viz.
\begin{equation}
    C_i[x,y] = A_i \exp{(-[x^2+y^2]{\sigma^{-2}})}
\end{equation}
\noindent where $\sigma$ is drawn from a uniform distribution between 0 and 1. We tested how (a) the presence/absence of a relationship between brightness $A_i$ and size $\sigma$, and (b) the presence/absence of a relationship between $A_i$ for different channels affects performance. When $A_0 \propto \sigma$ (linear relationship between size and brightness) and $A_i = A_0^i$ (i.e. relationship across channels), the CNN solution performed the best. When the inter-channel relationship was removed (i.e. each $A_i$ was randomly sampled), the performance suffered. Finally, when both $A_i$ and $\sigma_i$ were both independently randomly sampled for all channels, the model achieved a performance that is equivalent to randomly guessing. This is expected behaviour since $A_i$, $A_j$, $\sigma_i$, and $\sigma_j$ are uncorrelated for $i\ne j$.  These experiments suggest the CNN solution on SDOML images works (partly) because of inherent relationships between the morphology of solar structures and their brightness, and because of relationships between different wavelength channels. 

\section{Summary \& Future Work}
In this paper we have shown a CNN that takes multiple wavelengths as input and which auto-calibrates SDO/AIA data, that performs significantly above a non-CNN baseline (100\% vs. 54\%) for a tolerance level of 0.05.

This opens the doors for future HSO missions to auto-calibrate their E(UV) instruments without using sounding rockets, enabling deep space Sun observatories from many vantage points.

\subsubsection*{Acknowledgments}
This project was conducted during the 2019 NASA Frontier Development Lab (FDL) program, a public/private partnership between NASA, SETI and industry partners including Lockheed Martin, IBM, Google Cloud, Intel, and NVIDIA Corporation. The authors wish to thank in particular IBM for generously providing computing resources. We gratefully thank all our mentors for guidance and useful discussion, as well as the SETI Institute for their hospitality during the program.

\bibliography{autocalibration}

\end{document}